\documentclass{jfm}

\usepackage{graphicx}
\usepackage{fancyhdr}
\usepackage{epsfig}
\usepackage{rotating}
\usepackage{amssymb}
\usepackage{mathrsfs}
\usepackage[round]{natbib}
\usepackage{epsf}
\usepackage{amsmath}
\usepackage{color}
\usepackage{lscape}
\usepackage{setspace}
\usepackage{enumerate}
\usepackage{subcaption}
\usepackage[latin1]{inputenc}
\usepackage{textcomp}
\usepackage[english]{babel}
\usepackage[super]{nth}
\usepackage{bibentry}
\usepackage{trimclip}
\usepackage{xspace}

\newcommand{\f}{\frac}

\newcommand{\be}{\begin{equation}}
\newcommand{\ee}{\end{equation}}
\newcommand{\ba}{\begin{eqnarray}}
\newcommand{\ea}{\end{eqnarray}}
\newcommand{\md}{\mathrm{d}}
\newcommand{\mi}{\mathrm{i}}
\newcommand{\me}{\mathrm{e}}
\newcommand{\mF}{\mathcal{F}}

\DeclareMathOperator{\Lapl}{\mathcal{L}}

\def\Xint#1{\mathchoice
   {\XXint\displaystyle\textstyle{#1}}%
   {\XXint\textstyle\scriptstyle{#1}}%
   {\XXint\scriptstyle\scriptscriptstyle{#1}}%
   {\XXint\scriptscriptstyle\scriptscriptstyle{#1}}%
   \!\int}
\def\XXint#1#2#3{{\setbox0=\hbox{$#1{#2#3}{\int}$}
     \vcenter{\hbox{$#2#3$}}\kern-.5\wd0}}
\def\dashint{\Xint-}

\def\AR{\clipbox{0pt 0pt .32em 0pt}\AE\kern-.30emR}

\newcommand{\Ae}{\textit{Ae}\xspace}

\shorttitle{Aerodynamic theory of hovering membrane wings}
\shortauthor{S. Tiomkin and A. Gehrke}

\title{Unsteady aerodynamic theory and experiments of hovering membrane wings} 

\author{Sonya Tiomkin\aff{1}\corresp{\email{sonyat1@usf.edu}}  \and Alexander Gehrke\aff{2}}

\affiliation{\aff{1}Department of Mechanical and Aerospace Engineering, University of South Florida, Tampa,
FL 33620, USA

		\aff{2}Center of Fluid Mechanics, School of Engineering, Brown University, Providence,
RI 02912, USA}

\begin{document}

\maketitle

\begin{abstract}

We investigate the unsteady lift response of compliant membrane wings in hovering kinematics by combining analytical inviscid theory with experimental results. An unsteady aerodynamic model is derived for a compliant thin aerofoil immersed in incompressible inviscid flow of variable freestream velocity at high angles of attack. The model, representing a spanwise section of a hovering membrane wing, assumes small membrane deformation and attached flow. These assumptions are supported by experiments showing that passive membrane deformation suppresses flow separation when hovering at angles of attack up to $55^\circ$. 
An analytically derived expression is obtained for the unsteady lift response, incorporating the classical Wagner and Theodorsen functions and the membrane dynamic response. This theoretical expression is validated against experimental water-tank measurements that are performed on hovering membrane wings at angles of attack of $35^\circ$ and $55^\circ$. Data from membrane deformation measurements is applied to the theoretical lift expression, providing the theoretical lift response prediction for each of the available experimental scenarios.
Results of the comparison show that the proposed theory accurately predicts unsteady lift contributions from membrane deformation at high angles of attack, provided the deformation remains small and the flow is attached. 
This agreement between inviscid theory and experimental measurements suggests that when flow separation is suppressed, the unsteady aerodynamic theory is valid well beyond the typical low angle of attack regime.

\end{abstract}

%

\section{Introduction}

\begin{figure}
	\centering
	\includegraphics[]{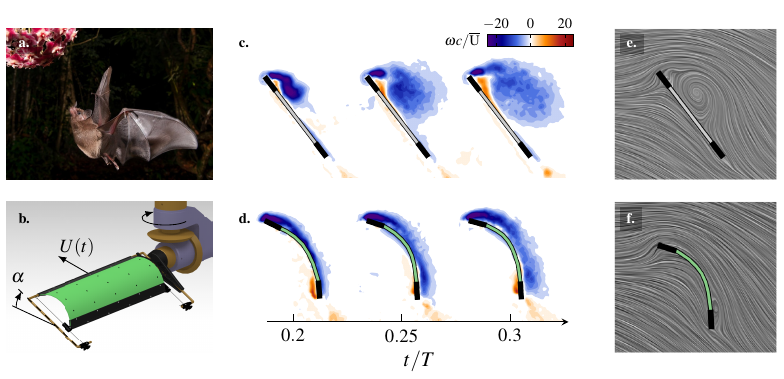}
	\caption{Experimental study and motivation:
		(a) Pallas's long-tongued bat (\textit{Glossophaga soricina}, photo by: Gregory Basco/www.deepgreenphotography.com),
            (b) Drawing of the experimental setup used by \cite{Gehrke_2022} in which the membrane wing is mounted on a flapping platform.
		Vorticity field snapshots of (c) rigid and (d) flexible membrane wings obtained for $\hat{\alpha}=55^\circ$ at three time instances, $t/T = 0.20, 0.25$, and $0.30$, where $T$ is the full-cycle time period that includes a forward and a backward stroke.  
		Streamlines around the (e) rigid and (f) flexible wing obtained at $t/T = 0.25$ demonstrate how passive membrane deformation reattaches the flow and reduces flow separation in hovering flapping motion at high angles of attack up to $55^\circ$.
	}
	\label{fig:Exp_model_intro}
\end{figure}

Flexible appendages such as fins and wings enable efficient and versatile locomotion in many animals.
In particular, many bat species, such as the Pallas's long-tongued bat (\textit{Glossophaga soricina}, figure~\ref{fig:Exp_model_intro}a) use thin compliant membrane wings to achieve efficient hovering and agile manoeuvring. 
These flexible wings offer aerodynamic advantages over rigid wings due to their passive deformation, which delays stall and allows for gust resilience, making them a promising model for bio-inspired flight technologies \citep[e.g.][]{Muijres2008,Chin2016}.

In hovering flight, natural fliers typically perform symmetric back-and-forth wing motions in a horizontal stroke-plane.
For rigid wings, these motions generate a leading-edge vortex that enhances lift but also increases drag and power demand.
Recent experiments on hovering membrane wings \citep{Gehrke_2022,gehrke_highly_2025} have shown that unsteady membrane deformation suppresses flow separation and yields up to a 20\% lift enhancement compared to rigid wings (figure~\ref{fig:Exp_model_intro}).

In a parallel theoretical effort to understand the unsteady aerodynamic performance of compliant membrane aerofoils, \citet{Tiomkin2022} extended the classical unsteady thin aerofoil theory of \cite{vonKarman1938} and \cite{Sears1940}, originally derived for a flat plate immersed in inviscid incompressible flow, to include the aeroelastic response of the membrane aerofoil to unsteady flow conditions due to a prescribed flapping motion or a transverse gust encounter. 
The extended theory provides closed-form expressions for the unsteady lift coefficient due to membrane deformation in inviscid incompressible unsteady flow assuming low angles of attack, small membrane camber, and a constant freestream velocity.
The analytical expressions yield the unsteady lift coefficient for any (small amplitude) membrane deformation in time, where the deformation can be obtained either by solving the membrane dynamic equation, as demonstrated by \cite{Tiomkin2022}, where simplifying assumptions such as constant tension along the membrane are generally required, or directly from experimental measurements. 

The current study aims to further extend the unsteady aerodynamic theory and connect between experimental measurements and theoretical predictions of the unsteady lift response of flexible membrane wings in hovering flight kinematics. 
To this end we develop the unsteady aerodynamic theory of compliant membrane aerofoils immersed in inviscid incompressible flow of variable freestream velocity at high angles of attack.
The analytical solution provides a novel closed-form expression for the unsteady lift coefficient due to membrane deformation of a surging membrane aerofoil at high angles of attack.
The theory is validated by incorporating experimental wing kinematics and membrane deformation data of \citet{Gehrke_2022} and \citet{gehrke_highly_2025} in the theoretical model and comparing the predictions to the measured lift coefficients for angles of attack of $35^\circ$ and $55^\circ$.

\section{Methodology}\label{sec:Methodology}

\subsection{Theoretical model}\label{sec:Formulation}

\begin{figure}
	\centering
    \includegraphics[width=0.9\textwidth]
    {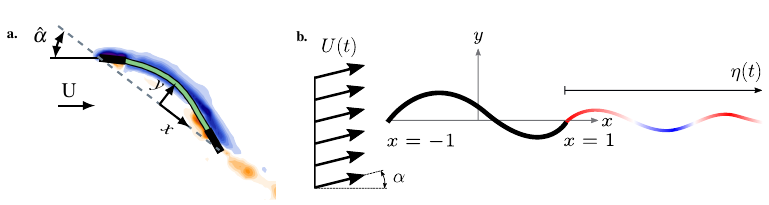}
	\caption{Representation of (a) the experimental membrane wing section and surrounding flowfield and (b) the theoretical model used to derive the unsteady lift coefficient acting on a flexible membrane aerofoil in variable freestream velocity. PIV results shown on the experimental model support the assumptions of attached flow and wake extension along the chord line, utilized in the theoretical model.}
	\label{fig:TheoreticalModel}
\end{figure}

We consider a thin membrane aerofoil that is held by supports at a distance $c=2b$ from the leading to the trailing edge. 
The membrane is immersed in a uniform inviscid incompressible flow of fluid density $\rho$ and variable speed $U(t)$ at an angle of attack $\alpha$ (see figure~\ref{fig:TheoreticalModel}b), as encountered by an aerofoil section along the span of a hovering membrane wing (figures~\ref{fig:Exp_model_intro}b and \ref{fig:TheoreticalModel}a). 
The unsteady lift acting on the membrane aerofoil is derived here assuming small membrane deformation $(|y_x|\ll 1)$ and attached flow at high angles of attack.
These assumptions are justified by the experimental results of \cite{gehrke_highly_2025}, in which unsteady membrane deformation was shown to suppress the formation of leading edge vortices and flow separation, even at angles of attack as high as $55^\circ$ for large parts of the flapping cycle (see figures~\ref{fig:Exp_model_intro}d and \ref{fig:Exp_model_intro}f).

Under the above assumptions, the normal velocity along the aerofoil due to membrane deformation (normalised by $U(t)$) becomes
\be
\label{eq:Wad}
w_{a_d}(\eta,x) = -\cos{\alpha}\, y_x (\eta,x) - y_\eta (\eta,x) , \qquad x\in \left(-1,1\right),
\ee
where $x$ is a coordinate along the chord, $y$ denotes the membrane profile, $\eta=\int_0^t V(t')\md t'$ is the distance travelled by the aerofoil in terms of semi-chord lengths, $V(t)=U(t)/\overline{U}$ is the nondimensional freestream velocity, and $\overline{U}$ is the mean freestream velocity. 
The above change of variables from the standard use of time $t$ to the travelling distance $\eta$, first introduced by \cite{wu1971_part1}, allows the derivation of closed-form expressions for the unsteady aerodynamic load and the resulting lift coefficient in cases of variable freestream velocity, as detailed below. 
Note that for a constant freestream velocity we obtain ${V(t)=1}$ and $\eta(t) = t$, and the generalised solution presented here will converge to the solution of \cite{Tiomkin2022} in cases of small angles of attack. 

The fundamental equation of thin aerofoil theory, expressed in terms of $\eta$, is given by 
\be
\label{eq:ThinAirfoil-Fundamental}
\f{1}{2\upi}\dashint_{-1}^1\f{\gamma(\eta,\xi)}{x-\xi}\md\xi= w_{a_d}(\eta,x) - \f{1}{2\upi}\int_{1}^{1+\eta}\f{\gamma_w\left(\eta,\xi\right)}{x-\xi}\md\xi, \qquad x\in \left(-1,1\right),
\ee
in which $b, b/\overline{U}$ and $U(t)$ are used as the units of length, time, and circulation per unit length, respectively. 
Here $\gamma$ is the normalised bound vorticity per unit length along the profile, the dashed integral denotes the Cauchy principal value, and $\gamma_w (\eta,\xi)$ describes the normalised vorticity per unit length at location $\xi$ along the wake, $\xi\in \left[1,1+\eta\right]$, at time $t$ when the aerofoil has travelled a total distance of $\eta(t)$ (see figure~\ref{fig:TheoreticalModel}b).

Wake vortices are assumed to be continuously shed from the trailing edge into a flat wake at the instantaneous freestream velocity with a fixed strength. Namely, the wake vorticity distribution, $\gamma_w (\eta,\xi)$, is equivalent to the vorticity at the trailing edge at an aerofoil displacement of $\eta-\xi+1$,
\be
\gamma_w (\eta,\xi)=\gamma_w (\eta-\xi+1,1)\triangleq \gamma_{_{T\!E}}(\eta-\xi+1).
\ee
The above assumptions are supported by experimental observations of the wake (e.g. figure~\ref{fig:TheoreticalModel}a), in which the wake extends along the aerofoil chord-line in the vicinity of the membrane trailing edge.
By following the standard application of S\"{o}hngen's inversion formula to \eqref{eq:ThinAirfoil-Fundamental} and enforcing Kelvin's theorem \citep[cf.,][p. 289]{Sohngen1939,Bisplinghoff_book1996} we obtain
\be
\label{eq:GammaInt-step5}
2\int_{-1}^{1} \sqrt{\f{1+\xi}{1-\xi}}\,w_{a_d}(\eta,\xi)\md\xi = -\int_{0}^{\eta}\sqrt{\f{\zeta+2}{\zeta}}\gamma_{_{T\!E}} (\eta-\zeta) \md\zeta .
\ee
\cite{Tiomkin2017} showed that for a constant freestream velocity, for which ${V(t)\equiv 1}$ and ${\eta\equiv t}$, application of the standard Laplace transform to \eqref{eq:GammaInt-step5} yields a closed-form expression for the wake vorticity distribution in the Laplace plane. In the current case, we utilise the Laplace transform suggested by \cite{wu1971_part1} to derive a similar closed-form expression for the wake vorticity in the variable velocity case, where the Laplace transform is defined as
\be
\label{eq:Laplace_def}
\bar{F}(s) = \Lapl \left\{F(\eta);s\right\} = \int_0^\infty F(\eta) \me^{-s\eta}\md\eta .
\ee

The solution is obtained by utilising a Fourier cosine series expansion to describe the membrane slope in terms of the aerofoil displacement, $\eta$, 
\be
\label{eq:SailSlope-Fourier_eta}
y_x(\eta,\theta)=\f{1}{2}\mF_0(\eta)+\sum_{n=1}^\infty \mF_n(\eta)\cos n\theta ,
\ee
where we apply the standard coordinate transformation, $x=-\cos\theta$, which places the aerofoil leading edge at $x=-1$ ($\theta=0$) and the trailing edge at $x=1$ ($\theta=\upi$).
Note that the odd-index Fourier coefficients denote here the amplitude of the corresponding membrane mode-shape in the decomposition of the membrane deformation, while a combination of the even-index Fourier coefficients provides the even mode-shapes \citep[see][]{Tiomkin2017}.
By substituting \eqref{eq:SailSlope-Fourier_eta} into the normal velocity expression \eqref{eq:Wad} and applying the Laplace transform to \eqref{eq:GammaInt-step5} we obtain an expression for the wake vorticity distribution in the Laplace domain, 
\be
\label{eq:gamma_TE-Lap}
\bar{\gamma}_{T\!E}(s)= -\f{2\upi \me^{-s}}{K_0(s)+K_1(s)} \bar{f}_m(s) = -2\upi\, \bar{\Psi}(s) s^2 \bar{f}_m(s) ,
\ee
where $\bar{\Psi}(s)$ is the Laplace transform of K\"{u}ssner's function \citep[e.g.,][]{Sears1940}, $K_0$ and $K_1$ are modified Bessel functions of the second kind, and $\bar{f}_m(s)$ is an auxiliary function in the Laplace domain given by
\ba
\nonumber
\bar{f}_m(s) &=& \cos{\alpha}\left\{\f{1}{2}\bar{\mF}_{1}(s) - \f{1}{2}\bar{\mF}_{0}(s)\right\}  -\f{1}{4} s\bar{\mF}_0(s) -\f{1}{4} s\bar{\mF}_1(s) + \f{1}{4} s\bar{\mF}_2(s)\\
&&+ \sum_{m=2}^{N/2}  \f{s\bar{\mF}_{2m-1}(s)}{(2m-1)^2-1} . \label{eq:fm_s}
\ea

The aerodynamic load along the aerofoil is then expressed by adapting the method of \cite{Schwarz1940} to the variable freestream velocity formulation, yielding
\ba
\label{eq:DP-GeneralSol}
\f{\Delta C_p(\eta,x)}{V(t)^2}&=& -\f{4}{\upi}\sqrt{\f{1-x}{1+x}}\dashint_{-1}^{1} \sqrt{\f{1+\xi}{1-\xi}}\,\f{w_{a_d}(\eta,\xi)}{x-\xi}\md\xi + \f{4}{\upi}\dashint_{-1}^{1}\Lambda_1 (x,\xi) \f{\p w_{a_d}}{\p\eta}\md\xi
\nonumber
\\
&&\mbox{} + \f{2}{\upi}\sqrt{\f{1-x}{1+x}} \int_{0}^{\eta} \f{\gamma_{_{T\!E}} (\eta-\zeta)}{\sqrt{\zeta (\zeta+2)}}\md\zeta , 
\ea
where $\Lambda_1$ is an auxiliary function given by
\be
\label{eq:Lam1}
\Lambda_1\left(x,\xi\right)=\ln\left|\f{\sqrt{(1-x)(1+\xi)}+\sqrt{(1+x)(1-\xi)}} {\sqrt{(1-x)(1+\xi)}-\sqrt{(1+x)(1-\xi)}}\right| .
\ee
It is convenient to denote the integral terms on the right hand side of \eqref{eq:DP-GeneralSol} by
$\widetilde{\Delta C}_{p_0}$, $\widetilde{\Delta C}_{p_1}$, and $\widetilde{\Delta C}_{p_2}$, which respectively represent the quasi-steady, apparent mass, and wake contributions to the aerodynamic load along the aerofoil.
Closed-form expression are readily obtained for these terms by substituting \eqref{eq:Wad} into the $\widetilde{\Delta C}_{p_0}$ and $\widetilde{\Delta C}_{p_1}$ terms, while the wake term, $\widetilde{\Delta C}_{p_2}$, is most conveniently expressed in the Laplace plane by applying the convolution theorem and utilising the wake vorticity expression given in the Laplace domain \eqref{eq:gamma_TE-Lap}. Inverse Laplace transform is then required to obtain the total aerodynamic load in time  (or $\eta$) domain.

For non-small angles of attack, as considered here, integration on the aerodynamic load \eqref{eq:DP-GeneralSol} along the membrane chord-line leads to the normal aerodynamic force coefficient due to membrane deformation, 
\be
\label{eq:Cn_d_U_varying-General}
C_{n_d}(t) = 2\upi V(t)^2 \left\{ \int_0^{\eta(t)} \Phi(\eta(t)-\tau)\dot{f}_m(\tau)\md\tau +g_m(\eta(t)) \right\}, 
\ee
where $\Phi(t)$ is the time-domain Wagner function, an upper dot represents a derivative with respect to $\eta$, and $f_m(\eta)$ and $g_m(\eta)$ are auxiliary functions given by 
\ba
\nonumber
f_m(\eta) &=& \cos{\alpha}\left\{\f{1}{2}\mF_{1}(\eta) - \f{1}{2}\mF_{0}(\eta)\right\}  -\f{1}{4} \dot{\mF}_0(\eta) -\f{1}{4} \dot{\mF}_1(\eta) + \f{1}{4} \dot{\mF}_2(\eta)\\
&&+ \sum_{m=2}^{N/2}  \f{\dot{\mF}_{2m-1}(\eta)}{(2m-1)^2-1}, \label{eq:fm_t}\\
\nonumber
g_m(\eta) &=&  \cos{\alpha}\left\{\f{1}{4}\dot{\mF}_2(\eta)-\f{1}{4}\dot{\mF}_0(\eta)\right\}  - \f{3}{16} \ddot{\mF}_1(\eta)  + \f{1}{8} \ddot{\mF}_3(\eta)  \\
&&+\f{1}{2} \sum_{m=3}^{N/2} \f{\ddot{\mF}_{2m-1}(\eta)}{(2m-1)^2-1}  ,
\label{eq:gm_t}
\ea
where $f_m(\eta)$ is the $\eta$-domain representation of \eqref{eq:fm_s}.
Note that $g_m(\eta)$ represents the normalised non-circulatory lift coefficient, while $f_m(\eta)$ describes the circulatory term, as it is directly related to the wake vorticity through \eqref{eq:gamma_TE-Lap}.
For a constant freestream at small angles of attack $(\cos{\alpha}\cong 1, \eta\equiv t)$ the solution described above in \eqref{eq:Cn_d_U_varying-General}-\eqref{eq:gm_t} coincides with the unsteady lift coefficient given by \cite{Tiomkin2022}.

In cases of harmonic aerofoil motions of reduced frequency $k$, the normal force coefficient in \eqref{eq:Cn_d_U_varying-General} becomes
\be
\label{eq:Cn_d_U_varying}
C_{n_d}(t) = 2\upi V(t)^2\left\{ C(k){f}_m\left(\eta \left(t\right)\right) +g_m\left(\eta \left(t\right)\right) \right\} ,
\ee
and the unsteady lift coefficient due to membrane deformation is 
\be
\label{eq:Cl_d}
C_{l_d}(t) = C_{n_d}\cos{\alpha} =  2\upi\cos{\alpha}\, V^2(t)\left\{ C(k){f}_m\left(\eta \left(t\right)\right) +g_m\left(\eta \left(t\right)\right) \right\} ,
\ee
where $C(k)$ is the frequency-domain Theodorsen's function,
\be
\label{eq:C_Th-FreqDomain}
C(k) = \f{H_1^{(2)}(k)}{H_1^{(2)}(k)+\mi H_0^{(2)}(k)},
\ee
and $H_0^{(2)}$ and $H_1^{(2)}$ are Hankel functions of the second kind.

We further note that in cases where the first mode is significantly more dominant than all other modes $(\mF_1\gg\mF_n\; \forall\, n\neq1)$, the membrane deformation can be represented by $y(\eta,\theta)=\f{1}{2}\mF_1(\eta) \sin^2{\theta}$, where $\f{1}{2}\mF_1(\eta)$ is the maximum camber of the aerofoil. Under these conditions a simplified expression is readily obtained for the lift coefficient that depends only on the wing kinematics and the first Fourier coefficient (or maximum membrane camber),
\be
\label{eq:Cl_d-approx}
C_{l_d}(t) \cong 2\upi\cos{\alpha}\, V^2(t)\left\{ C(k)\left[\f{1}{2}\mF_1(\eta)\cos{\alpha} -\f{1}{4}\dot{\mF}_1(\eta)\right] -\f{3}{16}\ddot{\mF}_1(\eta) \right\} .
\ee

The above expressions \eqref{eq:Cn_d_U_varying-General}-\eqref{eq:Cl_d} were derived assuming small membrane camber, $|y_x|\ll 1$, constant non-small angle of attack, and attached flow. For scenarios in which these assumptions apply, the unsteady lift coefficient produced by the membrane deformation is given by \eqref{eq:Cl_d} and is determined by the flow parameters ($\alpha$, $V(t)$, and $k$) and the Fourier coefficients that decompose the membrane deformation in time. While the flow parameters (or wing kinematics) are generally given for every scenario considered, the Fourier coefficients can be obtained either through a numerical solution of the coupled membrane-fluid problem or by using membrane deformation measurements to predict the lift.

In the current work we choose the latter approach and apply the Fourier series expansion \eqref{eq:SailSlope-Fourier_eta} to experimental data of the membrane deformation in time, from which time-dependent Fourier coefficients are computed (figure~\ref{fig:modelOverview}). 
These Fourier coefficients, $\mF_n(\eta)$, when substituted in \eqref{eq:fm_t}, \eqref{eq:gm_t}, and \eqref{eq:Cl_d} provide a theoretical prediction of the sectional lift coefficient due to membrane deformation in the experimental conditions. Therefore, this method loosens the requirements of a constant tension in the membrane and small angles of attack, as commonly assumed in theoretical studies, while still limiting the validity of the results to scenarios of small membrane deformations and attached flow.

\subsection{Experimental configuration}\label{sec:setup}

The experimental setup of \cite{Gehrke_2022} is briefly described here in the context of the current work, as this work serves for validation of the novel theoretical model presented in \S\ref{sec:Formulation}.
The experimental configuration features a membrane wing that is attached to a rigid frame that performs a prescribed flapping motion in a water tank.
The membrane wing includes rigid leading and trailing edges that are free to rotate, allowing the wing edges to align favourably with the flow during each flapping cycle.
An additional degree of freedom is maintained at the trailing edge by permitting free translation of the edge in the chord-wise direction, varying the membrane chord-length in time.

This configuration enables passive deformation of the thin aerofoil shape under aerodynamic loading, creating a natural synergy between structural flexibility and aerodynamic forces, characterised by the aeroelastic number, $\Ae={Eh}/({\f{1}{2}\rho  \overline{U}^2 c})$, where $E$ is the Young modulus of the membrane.
The model wing performs a sinusoidal stroke along the horizontal plane, and a trapezoidal angle of attack ($\alpha$) profile, similar to the hovering kinematics of small bats and insects (see figures~\ref{fig:Exp_model_intro}b and \ref{fig:modelOverview}b showing the experimental setup and wing kinematics).
The experimental study focused on capturing the membrane dynamic response and the unsteady aerodynamic forces produced by the flapping wing for a wide range of aeroelastic numbers, $\Ae = 0.25-12$, and angle-of-attack amplitudes of $\hat{\alpha}=15^\circ-75^\circ$. 
Aerodynamic forces and moments were recorded at the wing root with a 6-axis force/torque transducer, while the wing's deformation was captured throughout the flapping cycle with two machine vision cameras in stereo-configuration and reconstructed using photogrammetry.
Three-dimensional deformation results showed no significant spanwise variation in membrane profile, allowing its representation by a single two-dimensional profile \citep[][]{gehrke_highly_2025}.
Finally, the unsteady flowfield around the membrane wing was recorded using planar particle image velocimetry (PIV) at a fixed span-wise position, located at the radius of the second moment of area ($r = 0.55R$, where $R$ is the total wingspan).
More information about the experimental setup, data acquisition and processing, and in-depth aerodynamic analysis can be found in \cite{Gehrke_2022} and \cite{gehrke_highly_2025}.

\subsection{Connecting theory to measurements }

\begin{figure}
	\centering
	\includegraphics[width=\textwidth]{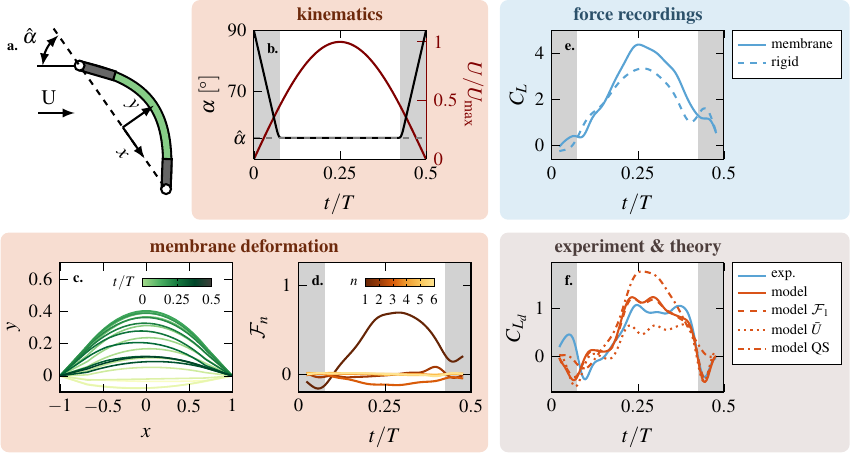}
	\caption{(a) Sketch of the flexible membrane wing profile,
            (b) Angle of attack ($\alpha$) and local wing velocity ($U$) profiles as a function of normalised time $t/T$,    
		(c) Membrane camberline, $y(x)$, obtained at different times ($t/T = 0$ to $0.5$),
		(d) Membrane Fourier coefficients, $\mF_n$, as a function of time $t/T$,
            (e) Measured lift coefficients, $C_L$, as a function of time $t/T$ for a membrane and a rigid wing,
		(f) Finite-wing lift coefficient due to membrane deformation, $C_{L_d}$, as a function of time $t/T$; comparing the experimental measurements with the theory prediction, along with simplified model predictions assuming single-mode oscillations ($\mF_1$), constant freestream velocity ($\overline{U}$), and a quasi-steady aerodynamic model (QS). Very good agreement is observed between the theoretical and measured lift coefficients, computed for $\hat{\alpha}=35^\circ, \Ae=2.5$, which is also captured by the single-mode approximation. However, the simplified constant $\overline{U}$ model and the quasi-steady model do not agree well with measurements.
		}
	\label{fig:modelOverview}
\end{figure}

Seeking to validate the unsteady aerodynamic theory derived in \S\ref{sec:Formulation}, membrane deformation measurements obtained by \cite{Gehrke_2022} along the mid-wingspan section are converted first to body-fixed coordinates, normalised by the instantaneous semi-chord length, and decomposed into a Fourier series expansion \eqref{eq:SailSlope-Fourier_eta} (see figures~\ref{fig:modelOverview}c and \ref{fig:modelOverview}d, respectively).
The resultant Fourier coefficients are then used to compute the theoretical sectional lift coefficient, $C_{l_d}(t)$, via \eqref{eq:Cl_d}. 
This result yields the finite-wing lift contribution from membrane deformation,
\be
\label{eq:CLd_finite_wing}
C_{L_d}(t) = \f{C_{L_\alpha}}{2\upi} \f{c(t)}{\overline{c}}  C_{l_d}(t) ,
\ee
where $c(t)$ and $\overline{c}$ are the instantaneous and reference chord-length, respectively.
The finite-wing lift slope $C_{L_\alpha}$ is obtained using the two-station approximation \citep[e.g.][pp. 451-454]{Anderson_book2024} for a rectangular wing of aspect ratio \AR$=2.7$, matching the wing geometry used in the experiments of \citet{Gehrke_2022}. 

The corresponding experimental value for the lift coefficient due to membrane deformation is obtained by calculating the difference between the lift coefficient measured for the membrane wing and the lift coefficient obtained on a rigid flat plate that undergoes an equivalent prescribed flapping motion (see figures~\ref{fig:modelOverview}e). The resultant experimental lift coefficient due to membrane deformation is then compared with the corresponding theoretical value~\eqref{eq:CLd_finite_wing} for every experimental scenario considered (e.g. figure~\ref{fig:modelOverview}f). 
Note that for cases of dominant first-mode oscillations the sectional lift coefficient in \eqref{eq:CLd_finite_wing} can be obtained from the approximate expression \eqref{eq:Cl_d-approx}. In the scenario presented in figure~\ref{fig:modelOverview}, the first Fourier coefficient is clearly the most dominant coefficient and therefore the lift coefficient predicted with the single mode approximation is practically indistinguishable from the full model prediction. 
However, for more general cases, in which higher mode oscillations are observable in the membrane dynamic response, the full model \eqref{eq:Cl_d} is necessary to predict the sectional lift coefficient.
Such higher-mode oscillations are expected, for example, in cases of higher reduced frequency, beyond the first natural frequency of the membrane.
An alternative use of the standard unsteady aerodynamic model with constant freestream velocity significantly under-predicts the lift coefficient, while a quasi-steady model strongly over-predicts the response (figure~\ref{fig:modelOverview}f), highlighting the importance of including the effect of the variable freestream velocity and the shedding wake in the predictive model.
Note that the gray areas in figures~\ref{fig:modelOverview}b and \ref{fig:modelOverview}d-f mark the time intervals during which the angle of attack varies in time in the experiments, violating the theory assumptions. Therefore, the theoretical prediction, in which only effects of the instantaneous angle of attack are considered, is expected to agree with the experimental results for $0.08<t/T<0.42$, with a less favourable agreement in the gray-shaded areas.

\section{Results and discussion}\label{sec:results}

\begin{figure}
	\centering
	\includegraphics[]{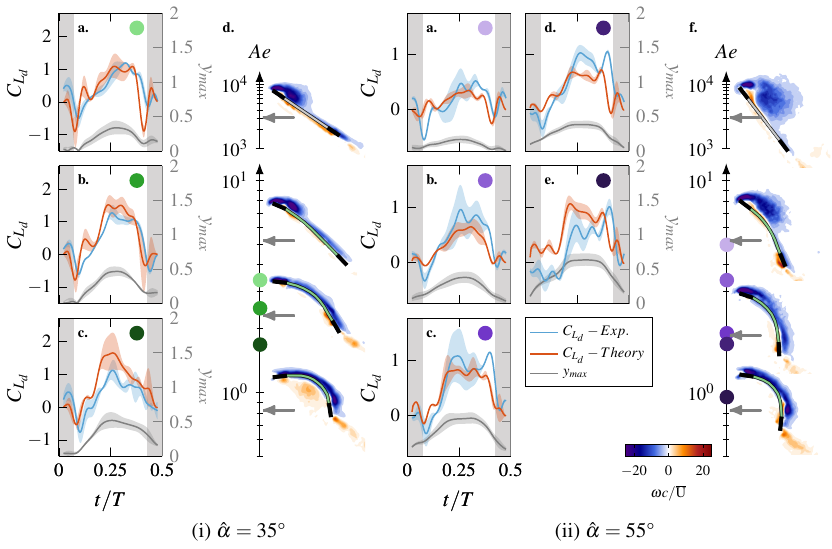}
	\caption{Lift coefficient, maximum membrane deformation, and flowfield snapshots obtained for varying values of aeroelastic number, $Ae$, and an angle-of-attack amplitude of (i) $\hat{\alpha}=35^\circ$; (ii) $\hat{\alpha}=55^\circ$. 
		Vorticity field snapshots in (i.d) and (ii.f) depict the flow around the rigid wing and flexible membrane wings at $t/T=0.25$ for (i.d) $\hat{\alpha}=35^\circ, \Ae=0.85,2.3,5.2$, and (ii.f) $\hat{\alpha}=55^\circ, \Ae=0.85,1.9,5.2$.
		Lift coefficient due to membrane deformation results in (i.a-c) and (ii.a-e) are computed for $\Ae=3.4,2.5,1.7$, and $\Ae=5,3.4,1.9,1.7,0.95$, respectively, with experimental measurements appearing in blue and theoretical prediction based on measured deformations in red.
		Black lines depict the measured maximum mean membrane camber, $y_{max}$.
		Shaded areas represent results obtained from forward and backward strokes and solid lines depict the mean values. Note that the mean freestream in the experimental setup is zero, so there is no practical difference between forward and backward strokes. Very good agreement is observed between theory and measurement for cases of $\Ae>1.7$. Lower values of \Ae result in high membrane camber that leads to flow separation near the trailing edge, as evident in the flowfield snapshots.  
	}
	\label{fig:CLd_ThExp_all}
\end{figure}

The unsteady lift coefficient of a finite membrane wing undergoing a prescribed flapping motion is studied for angle-of-attack amplitudes of $\hat{\alpha}=35^\circ$ and $55^\circ$ with various values of the aeroelastic number, as available from experimental data of \cite{Gehrke_2022} (see figures~\ref{fig:CLd_ThExp_all}i.a-c and \ref{fig:CLd_ThExp_all}ii.a-e, respectively).
For each of the available scenarios, determined by $\hat{\alpha}$ and \Ae, the experimental and theoretical lift coefficient results, and the maximum measured deformation are illustrated with blue-, red- and gray-shaded areas, respectively.
The shaded regions show data from the forward and backward strokes, while the solid lines indicate their average.
Since the wing flaps in a still-water tank, there is no imposed asymmetry between stroke directions.
However, minor variations in membrane deformation arise between the two strokes due to experimental variability.
Accordingly, we treat them as separate realizations of the same scenario (characterised by $\hat{\alpha}$ and \Ae) when applying the aeroelastic model.

For an angle-of-attack amplitude of $\hat{\alpha}=35^\circ$ very good agreement is observed between the theoretical prediction and the load-cell measurements obtained for moderate and high \Ae values of $\Ae=3.4$ and $2.5$ (figures~\ref{fig:CLd_ThExp_all}i.a and \ref{fig:CLd_ThExp_all}i.b, respectively). However, as the aeroelastic number is decreased to $1.7$, the maximum membrane camber increases to $26\%$ of the chord-length at mid-stroke and the theory overestimates the lift coefficient amplitude compared to the measured value. 
This is typical of cases in which trailing-edge separation appears on highly cambered aerofoils, as is indeed seen in the PIV results of the low-\Ae membrane (figure~\ref{fig:CLd_ThExp_all}i.d).
Notably, the general trend of the unsteady lift coefficient is captured well by the theory for all \Ae values examined here for $\hat{\alpha}=35^\circ$, and is strongly correlated to the maximum membrane camber. 

The difference between the lift coefficient that is measured during the forward and backward strokes of the wing corresponds to the respective change in maximum camber for $\hat{\alpha}=35^\circ$ and $\Ae>1.7$. However, for $\Ae=1.7$ this strong correlation between the variation in maximum camber and the measured lift coefficient is no longer apparent, especially in the initial acceleration stage $(t/T<0.15)$. At that time, the membrane maximum camber practically does not vary between forward and backward strokes, while a significant change is apparent in the measured lift coefficient, indicating the existence of other flow-induced effects that are not accounted for in the current theory. 

Note that as we compare here results of the lift coefficient due to membrane deformation, the model assumptions need to hold with regards to the difference between the flowfield obtained for a rigid wing and the flowfield about a compliant membrane wing, where both wings perform the same kinematic motion. Therefore, cases in which the flowfield around the membrane wing is very close to the one obtained for the rigid wing (e.g. second row in \ref{fig:CLd_ThExp_all}i.d and \ref{fig:CLd_ThExp_all}ii.f depicting results for $\hat{\alpha}=35^\circ,\Ae=5.2$ and $\hat{\alpha}=55^\circ,\Ae=5$, respectively) produce very good agreement between the two methods.

For an angle-of-attack amplitude of $\hat{\alpha}=55^\circ$ we again see good agreement between the theoretical prediction and measurements of the unsteady lift coefficient due to membrane deformation obtained for aeroelastic numbers between $1.9-5$, whereas values of $\Ae\le 1.7$ present significant differences between theoretical and experimental results ({figure~\ref{fig:CLd_ThExp_all}ii.a-e}).
For the lowest value of $\Ae=0.95$ the theory overestimates the lift coefficient due to an excessive membrane camber of more than $30\%$, whereas for $\Ae=1.7$ a moderate maximum camber of $20\%$ is obtained and the predictive tool underestimates the lift coefficient.
Flowfield snapshots for $\Ae=0.85$ and $\Ae=1.9$ (figure~\ref{fig:CLd_ThExp_all}ii.f, fourth and third lines, respectively) show that, within this range of the aeroelastic number, vortices are shed on the suction side near the trailing edge of the aerofoil, remaining attached for ${\Ae=1.9}$ and detaching for lower \Ae.
Accordingly, the theoretical model exhibits reduced accuracy at low aeroelastic numbers ($\Ae \le 1.7$) due to the observed flow separation.

\begin{figure}
	\centering
	\includegraphics[]{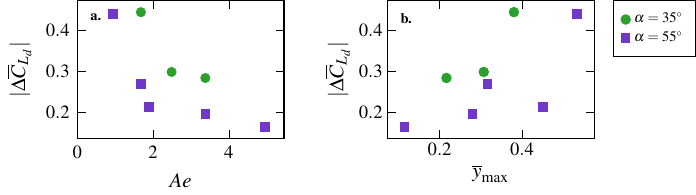}
  	\caption{Absolute value of the time-averaged error in prediction of the lift coefficient due to membrane deformation, $|\Delta \overline{C}_{L_d}|$, plotted against (a) aeroelastic number, \Ae; (b) mean maximum camber, $\overline{y}_{max}$, for angle-of-attack amplitudes of $\hat{\alpha}=35^o, 55^o$. Small \Ae values lead to considerable error in $C_{L_d}$, where $|\Delta \overline{C}_{L_d}|>0.4$, and as \Ae increases the error decreases.
	}
	\label{fig:CLd_Error}
\end{figure}

The time-averaged error between the predicted and measured lift coefficient due to membrane deformation is shown in figure~\ref{fig:CLd_Error} for all cases considered. Consistent with the time-dependent results in figure~\ref{fig:CLd_ThExp_all}, the cases of ${\Ae=1.7,\hat{\alpha}=35^o}$ and ${\Ae=1,\hat{\alpha}=55^o}$ exhibit the largest mean error, $|\Delta \overline{C}_{L_d}|\cong 0.44$, with corresponding mean maximum cambers of $19\%$ and $27\%$ of the chord, respectively. As $\Ae$ increases, the prediction error rapidly decreases and remains within $|\Delta \overline{C}_{L_d}| < 0.3$, consistent with the good qualitative agreement observed in figure~\ref{fig:CLd_ThExp_all}.
	
In general, low-error cases correspond to scenarios where the time-averaged maximum camber remains below $16\%$ $(\overline{y}_{\max}\le 0.32)$. An exception is the case of ${\Ae=1.9,\hat{\alpha}=55^o}$, which shows good agreement $(\Delta\overline{C}_{L_d}=0.21)$ despite a comparatively large camber $(\overline{y}_{\max}=0.45)$. Inspection of the flowfield for this case (figure~\ref{fig:CLd_ThExp_all}ii.f) indicates that the shear layer reattaches over most of the suction side, resulting in predominantly attached flow.
We therefore conclude that the primary requirement for accurate lift prediction using the present theoretical model is attached flow near the trailing edge. The small-camber condition, while related to maintaining attached flow, plays a secondary role.

\section{Concluding remarks}\label{sec:conclusions}

The unsteady aerodynamic theory of compliant aerofoils is extended to include effects of variable freestream velocity and high angles of attack, assuming small camber and fully attached flow. 
This scenario represents an aerofoil section along a membrane wing that performs a rotational flapping motion in a horizontal plane, inspired by the hovering flight of bats and motivated by the experimental work of \cite{gehrke_highly_2025} who showed that the flow about a hovering membrane wing can remain attached at high angle-of-attack amplitudes up to ${\hat{\alpha}=55^\circ}$.
The theory provides a closed-form expression for the unsteady lift coefficient due to membrane deformation, which depends on known functions (Theodorsen's and Wagner's functions) and the Fourier coefficients that decompose the membrane deformation in time. We further demonstrate how this expression can be applied to experimental deformation measurements and validate the predictive theory via comparison to the lift response measured by \cite{Gehrke_2022}.

We find very good agreement between the inviscid theory and the experimental results for aeroelastic numbers that are greater than $1.7$ with an angle-of-attack amplitude of $\hat{\alpha}=35^\circ, 55^\circ$. Lower values of \Ae result with an excessive membrane camber, which leads to vortex formation and flow detachment over the aerofoil. These flow and membrane characteristics violate the assumptions of the theoretical model, and therefore deteriorate the accuracy of the predictive theory.
Importantly, we find that cases in which the flow remains attached while the maximum camber is substantial still provide good agreement between theory and measurements, indicating the predominant role of the attached flow requirement compared to the small camber condition applied in potential theory.

In general, the good agreement obtained between analytical inviscid theory and measurements of the unsteady lift response of a hovering membrane wing in extreme conditions (high angles of attack) provides a promising validation of theory.
In all cases of attached flow and relatively small camber, the unsteady aerodynamic theory provides an accurate prediction of the unsteady lift (mean prediction error between $\Delta \overline{C}_{L_d} = 0.17$ and $0.27$), and can be used to compute the lift coefficient from any experimental or computational data of aerofoil deformation.

\section*{Acknowledgement}
The experimental data used in this study were collected by A.G. during his PhD at EPFL under the supervision of Prof. Karen Mulleners.
We are grateful to Prof. Mulleners for her valuable discussions and insights that helped shape the development of this work.

\vspace{0.2cm}
\noindent{\small \textbf{Decleration of Interests.} The authors report no conflict of interest.}


\bibliographystyle{jfm}
\bibliography{Bibliography}

\end{document}